\documentclass[twocolumn,showpacs,aps,prl,superscriptaddress]{revtex4-1}
\usepackage{graphicx,amssymb}
\usepackage{xcolor}
\pdfoutput=1

\newcommand{\sub}[1]{_{\mbox{\scriptsize{#1}}}}  

\begin{document}

\title{Stable spin domains in a non-degenerate ultra-cold gas}
\author{S.D.~Graham}
\affiliation{Department of Physics, Simon Fraser University, Burnaby, British Columbia V5A 1S6, Canada}
\author{D.~Niroomand}
\affiliation{Department of Physics, Simon Fraser University, Burnaby, British Columbia V5A 1S6, Canada}
\author{R.J.~Ragan}
\affiliation{Department of Physics, University of Wisconsin -- La Crosse, La Crosse, WI 54601, USA}
\author{J.M.~McGuirk}
\affiliation{Department of Physics, Simon Fraser University, Burnaby, British Columbia V5A 1S6, Canada}
\date{\today}

\begin{abstract}
We study the stability of two-domain spin structures in an ultra-cold gas of magnetically trapped $^{87}$Rb atoms above quantum degeneracy. Adding a small effective magnetic field gradient stabilizes the domains via coherent collective spin rotation effects, despite negligibly perturbing the potential energy relative to the thermal energy. We demonstrate that domain stabilization is accomplished through decoupling the dynamics of longitudinal magnetization, which remains in time-independent domains, from transverse magnetization, which undergoes a purely transverse spin wave trapped within the domain wall. We explore the effect of temperature and density on the steady-state domains, and compare our results to a hydrodynamic solution to a quantum Boltzmann equation.
\end{abstract}


\maketitle
Atomic Bose-Einstein condensates can form stable magnetic domains, for instance due to spin-dependent interaction energies \cite{hall1998domains,ketterle1998domains}, yet at higher temperatures diffusion of magnetic inhomogeneities in a weakly interacting non-degenerate gas is typically a fast process. So long as external field inhomogeneities are small enough that any differential spin-dependent energy is much less than the thermal energy, entropy dominates and the system rapidly relaxes to a near-uniform equilibrium. However, the addition of quantum coherence can inhibit diffusion and lead to stable, time-independent magnetic states, even above quantum degeneracy.

Macroscopic collective behavior in non-degenerate gases was observed in diffusion experiments with hydrogen \cite{lee1984Hspinwave} and helium \cite{nunes1992spin} that showed that connecting two reservoirs of spin-polarized gas led to unexpected nuclear magnetic resonance signals featuring coherence collapses and long-lived revivals. These results were interpreted as an instability-driven creation of domains \cite{castaing1984polarized}, which were the result of collective behavior induced by coherent exchange scattering in binary collisions, known as the identical spin rotation effect (ISRE) \cite{bashkin1981spin,lhuillier1982transport} -- similar to the Leggett-Rice effect \cite{Leggett1968,fuchs2003spinwaves}. Theoretical analysis suggested that the presence of a small magnetic field gradient allowed the existence of steady-state spin domains, giving rise to the surprisingly long-lived echoes \cite{fomin1994domains,ragan1997castaing,meyerovich1997relaxation}. However, direct imaging of the stable domains was elusive.

Collective effects with similar origins were seen in the hydrodynamic behavior of ultra-cold gases, beginning with the observation of spin waves in non-degenerate trapped Rb and K atoms \cite{mcguirk2002sw,thomas2008sw}. In addition to driving precessing spin currents that lead to coherent spin oscillations, exchange-driven collective behavior also drives collapse and revival of coherence in trapped ultra-cold gases \cite{mcguirk2011localized}, and can even be used to prolong coherence to extreme times \cite{reichel2010coherence}. The diffusion of spin inhomogeneities in ultra-cold gases is also strongly affected by exchange-driven collective behavior, including quantum-limited diffusion in strongly interacting Fermi gases \cite{sommer2011universal,kohl2013,trotzky2014observation} and significant deviations from classical diffusion in a weakly interacting Bose gas \cite{niroomand2015diffusion}.

Collective spin effects are particularly striking in weakly interacting gases because of the disparity of energy scales involved. Typical thermal energies of a non-degenerate ultra-cold gas are $k\sub{B}T/h \sim 15$~kHz, while mean-field interaction energies are a thousand times less. Instead, the requirement for significant collective spin effects is that the spin rotation parameter $\mu$ be large, where $\mu = \omega\sub{ex} \tau$, for elastic collision time $\tau$ and mean-field exchange frequency $\omega\sub{ex} = gn/\hbar$. The coupling constant $g=4\pi\hbar^2a/m$, with $s$-wave scattering length $a$, mass $m$, and density $n$. When $\mu$ is large, even small potential energy inhomogeneities can drive ensemble-wide dynamics, and the addition of small external effective magnetic fields can induce dramatic effects on the diffusive dynamics of such a system.

Earlier experiments in spin-polarized H and He were typically performed well into the hydrodynamic regime, where damping of small perturbations is often long, and thus spontaneously formed domain structures have time to emerge. This is not the case in the work presented here; instead, the system is initialized in a domain structure, and the stability of the state is explored. In this Letter, we report the existence of stable spin domains in a trapped non-degenerate gas of weakly interacting $^{87}$Rb atoms. By applying a small effective magnetic field gradient, the diffusion of magnetic inhomogeneities can be dramatically slowed and even stopped. We present direct evidence of spin domains persisting for more than 600~ms in regions where classical diffusion predicts relaxation timescales of less than 25~ms.

The experimental system consists of $^{87}$Rb atoms cooled to near degeneracy in an axisymmetric quasi-1D harmonic magnetic trap ($\omega_{z,\rho}=2\pi\times6.7, 255$~Hz respectively). The spinor is a pseudo-spin-1/2 doublet comprised of two hyperfine ground states coupled via a two-photon microwave transition at 6.8~GHz ($|1\rangle\equiv|F, m\sub{F}=1,-1\rangle$ and $|2\rangle\equiv|2,1\rangle$). A two-domain spin structure is initialized from $|1\rangle$ using an optical masking technique, wherein a masked off-resonant laser shifts one side of the atomic distribution out of resonance for a $\pi$-pulse, thereby transferring only the unmasked half of the distribution from $|1\rangle$ to $|2\rangle$ (see \cite{niroomand2015diffusion}). This coherent preparation maintains magnetization $|\vec{M}|=1$ throughout the ensemble.

The effective magnetic field is created with an optical dipole potential using a frequency- and amplitude-modulated acousto-optic modulator to produce a time-averaged linear gradient in optical intensity along the trap axis \cite{mcguirk2010optical}. A laser detuned $\sim$3.4~GHz from the D$_2$ excited state transition creates a differential energy shift that locally alters Larmor precession, analogous to a magnetic field gradient torquing real spins. The gradient size $G$ is adjusted with laser intensity. In all experiments, loss from spontaneous emission is below other loss processes, such as dipolar relaxation in $|2\rangle$-$|2\rangle$ collisions.

\begin{figure}[ht]
    \includegraphics[width=\linewidth,clip]{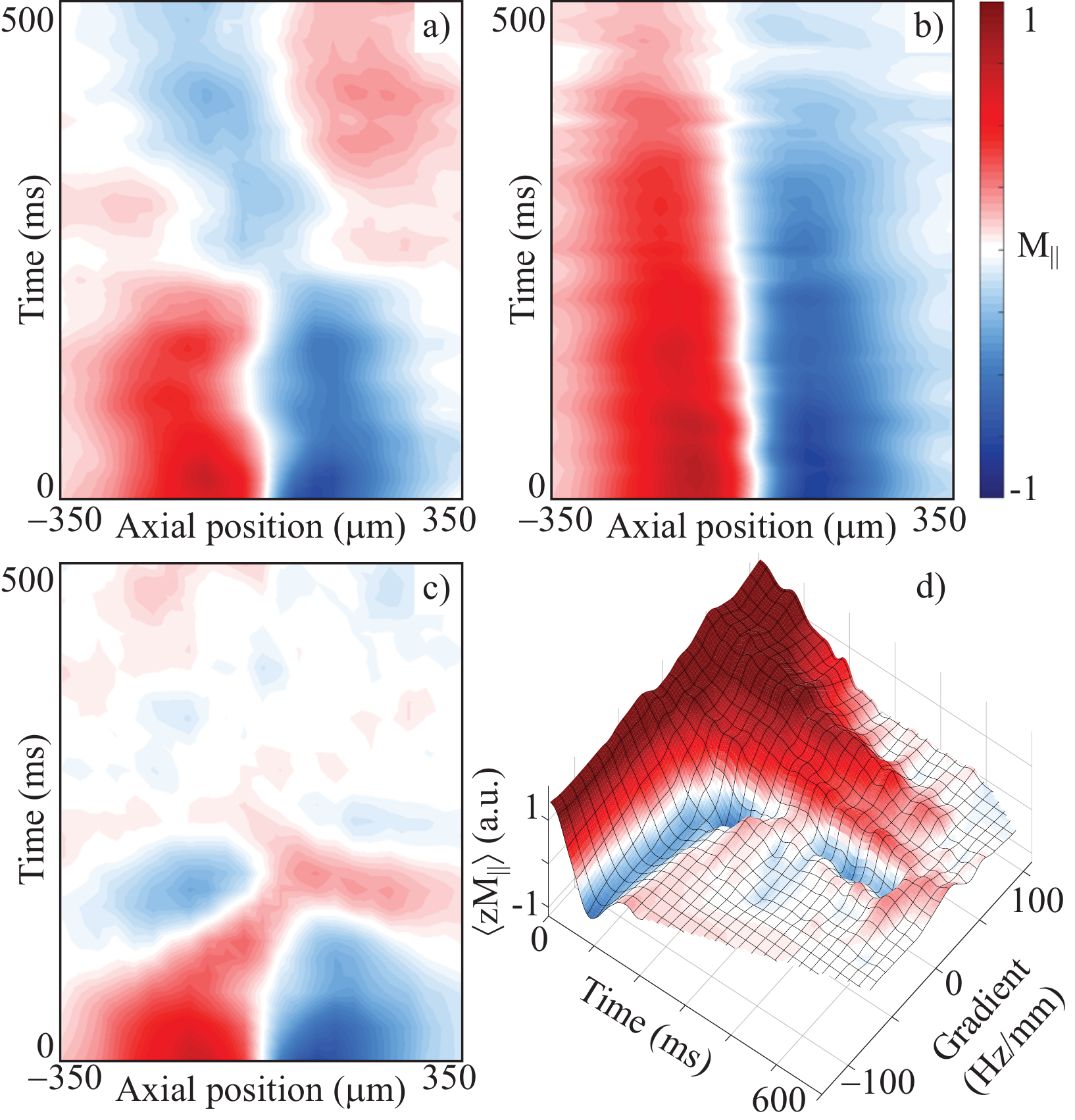}
    \caption{Spatiotemporal evolution of radially averaged $M_\parallel$ for $T=650$~nK and peak density $n_0=1.4\times10^{13}$~cm$^{-3}$ at $G =$ 0~Hz/mm (a), 54~Hz/mm (b), and -19~Hz/mm (c). The initial domain undergoes dipole oscillations at $G=0$, which increase in frequency and damping rate for $G<0$. Positive gradients stabilize the initial domain wall configuration. The domain wall drifts toward the $M_\parallel=1$ domain over time, as $|2\rangle$ decays via dipolar relaxation. d) Time evolution of the spin dipole moment, $\langle zM_\parallel \rangle$, in different effective field gradients. $\langle zM_\parallel \rangle$ oscillates for $G\leq0$, but stabilizes for positive gradients. Larger $G$ also leads to faster damping, through decoherence from the inhomogeneous applied field. }
    \label{Fig:profiles}
\end{figure}

Following preparation and evolution of the spin domains, the longitudinal magnetization is measured destructively by direct measurement of the populations of each spin state, $M_\parallel(z,t) = N_2(z,t) - N_1(z,t)$. Earlier work showed that, with no applied external field, domain structures in a uniform differential potential still undergo trap oscillations and diffusion [Fig.~\ref{Fig:profiles}(a)], despite significant slowing of dynamics due to collective behavior driven by coherent spin currents \cite{niroomand2015diffusion}. Fig.~\ref{Fig:profiles}(b) shows that addition of a small effective field gradient ($G=54$~Hz/mm) stabilizes the domains against both diffusion and trap oscillations.

Classically, diffusion should occur on a timescale given by the elastic collision time $\tau$, which at $\sim$24~ms for the conditions shown in Fig.~\ref{Fig:profiles} is much too fast to explain the domain lifetimes. Nor is the domain stabilization due to any differential mechanical force. For instance, a 50~Hz/mm field gradient shifts the relative trap centers by only $0.1~\mu$m, less than 0.1\% of the Gaussian half-width of the distribution $z_0$. This effect is instead driven by coherent spin interactions.

Furthermore, the sign of the effective field gradient is greatly important. Though the transport equation (Eq.~\ref{eq:Boltz} below) supports steady-state solutions for both positive and negative gradients, the initial domain preparation breaks this symmetry and leads to a preferred orientation. The sign of $G$ must oppose the initial spin gradient in the domain wall for the stable configuration to emerge. To rotate smoothly from one domain to the other as a spin traverses the domain wall, spin rotation from the external gradient must balance mean-field-induced spin torque. Otherwise, the system is initialized far from the gradient-driven stable state, and large amplitude transients consume virtually all the magnetization [Fig.~\ref{Fig:profiles}(c)]. The domain behavior as a function of $G$ can be summarized by dipole moment oscillations [Fig.~\ref{Fig:profiles}(d)].

The system is described theoretically by considering a quasi-one dimensional trapped gas with spin density distribution function $\vec{m}(z,p,t)$. The evolution of the magnetization in the presence of weak interactions is described by a 1D Boltzmann equation:
\begin{equation}\label{eq:Boltz}
  \partial_t \vec{m} + \partial_0 \vec{m}-\frac{1}{\hbar}\left(U\sub{diff}\hat{z} +g\vec{M} \right) \times \vec{m} = \partial_t \vec{m}|\sub{coll}
\end{equation}
where $\partial_0 = \frac{p}{m\sub{Rb}} \partial_z - m\sub{Rb} \omega_z^2 z \partial_p$, for mass $m\sub{Rb}$ and radially averaged collisional relaxation rate $\partial_t \vec{m}|\sub{coll}$ \cite{nikuni2002linear}. $\vec{M}(z,t) = \int \vec{m}\,dp/2\pi\hbar$ is the spatial component of the spin distribution, and is the observable in these experiments. $U\sub{diff}$ represents any differential energies experienced by the spin components and acts as an effective externally applied magnetic field.

Analysis of the kinetic equation in the presence of a linear effective magnetic field gradient, $U\sub{diff}/\hbar = G z$, indeed reveals the existence of steady-state solutions featuring two oppositely oriented longitudinal spin domains ($M_\parallel = \pm1$) joined by a narrow domain wall \cite{fomin1994domains,ragan1997castaing,meyerovich1997relaxation}. In the collisionless limit, Fomin derived a relation between the applied field gradient $G$ and the equilibrium size of the domain wall $\lambda\sub{eq}$ in an open normal Fermi liquid \cite{fomin1994domains}. Following these predictions, Ref.~\cite{ragan2004leggett} obtained a similar expression for a trapped gas in the hydrodynamic limit,
\begin{equation}
  G\sub{hydro} = \frac{\omega_z}{z_0} \frac{1}{\mu M} \left(\frac{\pi/2}{1.1\lambda\sub{eq}/z_0}\right)^3 \omega_z\tau, \label{eq:Gopt}
\end{equation}
which scales as $T/n$. To provide the spin rotation required to maintain stable domains, the domain wall must be nearly fully polarized -- that is, a helical domain wall where the spin smoothly rotates from one longitudinal orientation to another and $|\vec{M}|\simeq1$ -- so that $\mu M\gg1$.

\begin{figure}[ht]
    \includegraphics[width=\linewidth,clip]{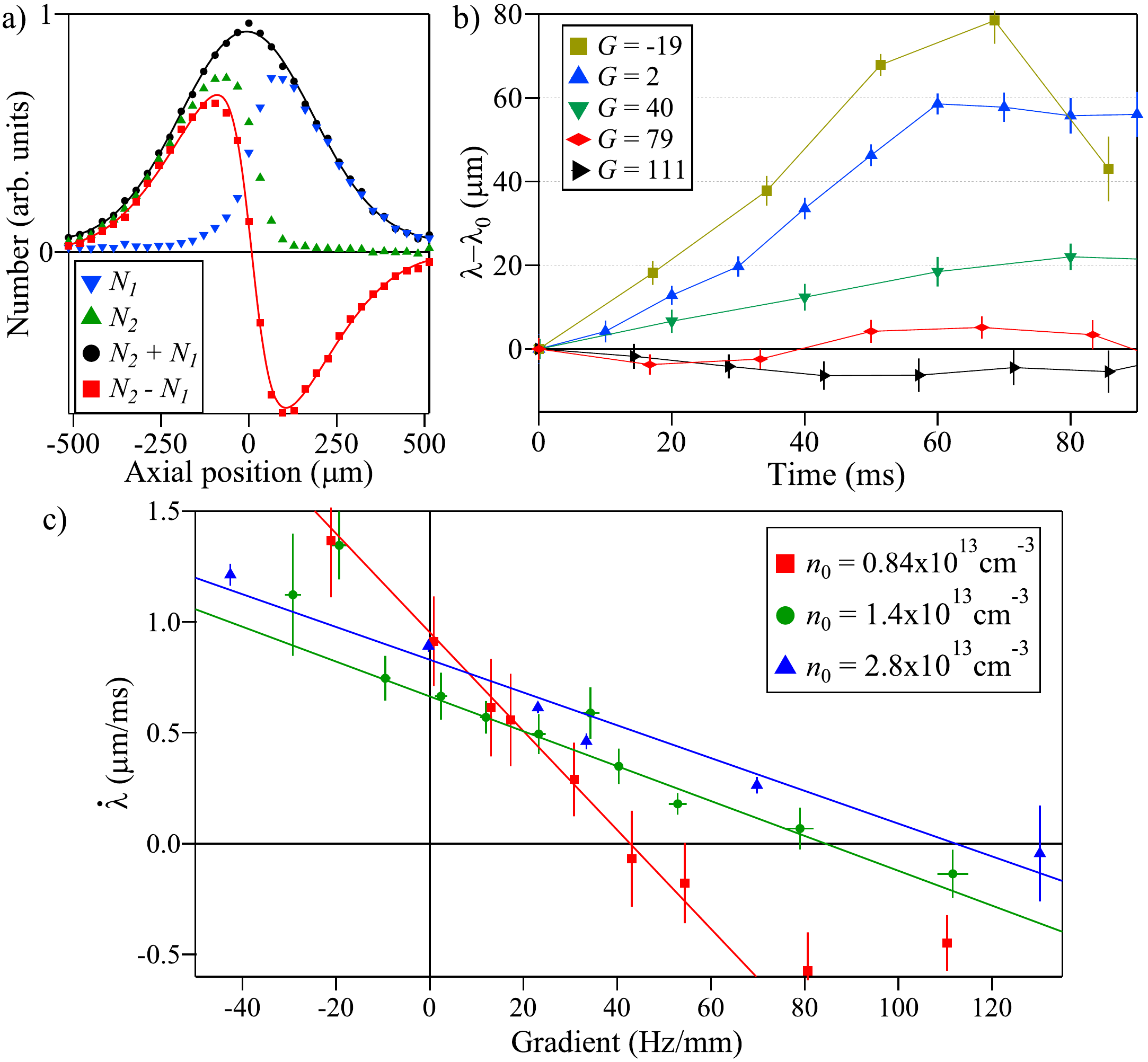}
    \caption{a) Normalized distribution of atoms in $|1\rangle$ ($N_1(z)$, $\blacktriangledown$) and $|2\rangle$ ($N_2(z)$, $\blacktriangle$) at $T=650$~nK and $n_0=2.8\times10^{13}$~cm$^{-3}$. A Gaussian is fit to the sum ($\bullet$) to extract temperature. The difference $M_\parallel = N_2-N_1$ (${\blacksquare}$) is fit with the same Gaussian multiplied by $\tanh z/\lambda$ to determine domain wall width $\lambda$. b) Time dependence of $\lambda$ immediately after application of $G$ (given in Hz/mm), exhibiting linear relaxation at short times. c) Initial domain-wall relaxation rate $\dot{\lambda}$ for three densities as a function of effective field gradient $G$. Uncertainties represent statistical uncertainty in fitting $\dot{\lambda}$ due to shot-to-shot temperature and density fluctuations. A linear fit is performed to determine the gradient $G_0$ that stabilizes the initial domain wall $\lambda_0$.}
    \label{Fig:lambda}
\end{figure}

To test these predictions, we study the dynamics of the domain wall, which provides insight into the behavior of the steady-state domains. We characterize the domain wall with a fit to a phenomenological model, $M_\parallel(z,t)=\exp(-z^2/2z_0^2)\tanh z/\lambda(t)$ , giving an initial domain wall width $\lambda_0 \simeq 73(3)~\mu$m [Fig.~\ref{Fig:lambda}(a)]. Changes in $\lambda$ due to the effective field gradient allow determination of the relationship between $G$ and $\lambda\sub{eq}$. When the applied gradient is small, the stable domain wall width $\lambda\sub{eq}$ is large -- much larger than $\lambda_0$ -- and the domain wall relaxes until it matches $\lambda\sub{eq}$ for that $G$. Conversely, if $G$ is large, then $\lambda$ shrinks. However, if $G$ is chosen so that $\lambda\sub{eq} = \lambda_0$, no domain-wall dynamics are observed, until eventually, at long times, dephasing from the effective field inhomogeneity removes enough coherence so that $\mu M <1$ and classical diffusion dominates.

We use these domain-wall dynamics to study the steady-state configuration, because one cannot wait until $\lambda$ has stabilized to measure $\lambda\sub{eq}$ as a function of $G$. Significant transients are observed during relaxation to the steady state, and damping of those transients removes magnetization, as does decoherence driven by the inhomogeneous effective field. Thus, by the time equilibration occurs, $|\vec{M}| \ll 1$, which in turn changes $\lambda\sub{eq}$. Instead, we focus on times much shorter than these damping timescales and determine which gradient $G_0$ stabilizes $\lambda_0$ in a fully polarized system.

Figure~\ref{Fig:lambda}(b) shows the initial behavior of $\lambda(t)$ for several gradients, where again $\lambda(t)$ comes from phenomenological fits to $M_\parallel(z,t)$. At short times, $\lambda(t)$ is approximately linear, and the rate of relaxation $\dot{\lambda}$ depends on the difference between $\lambda_0$ and $\lambda\sub{eq}$. Figure~\ref{Fig:lambda}(c) displays $\dot{\lambda}$ versus $G$; the horizontal intercept gives the gradient $G_0$ that produces the stable domain solution where $\lambda\sub{eq} = \lambda_0$ and $\dot{\lambda}=0$.

\begin{figure}[ht]
    \includegraphics[width=\linewidth,clip]{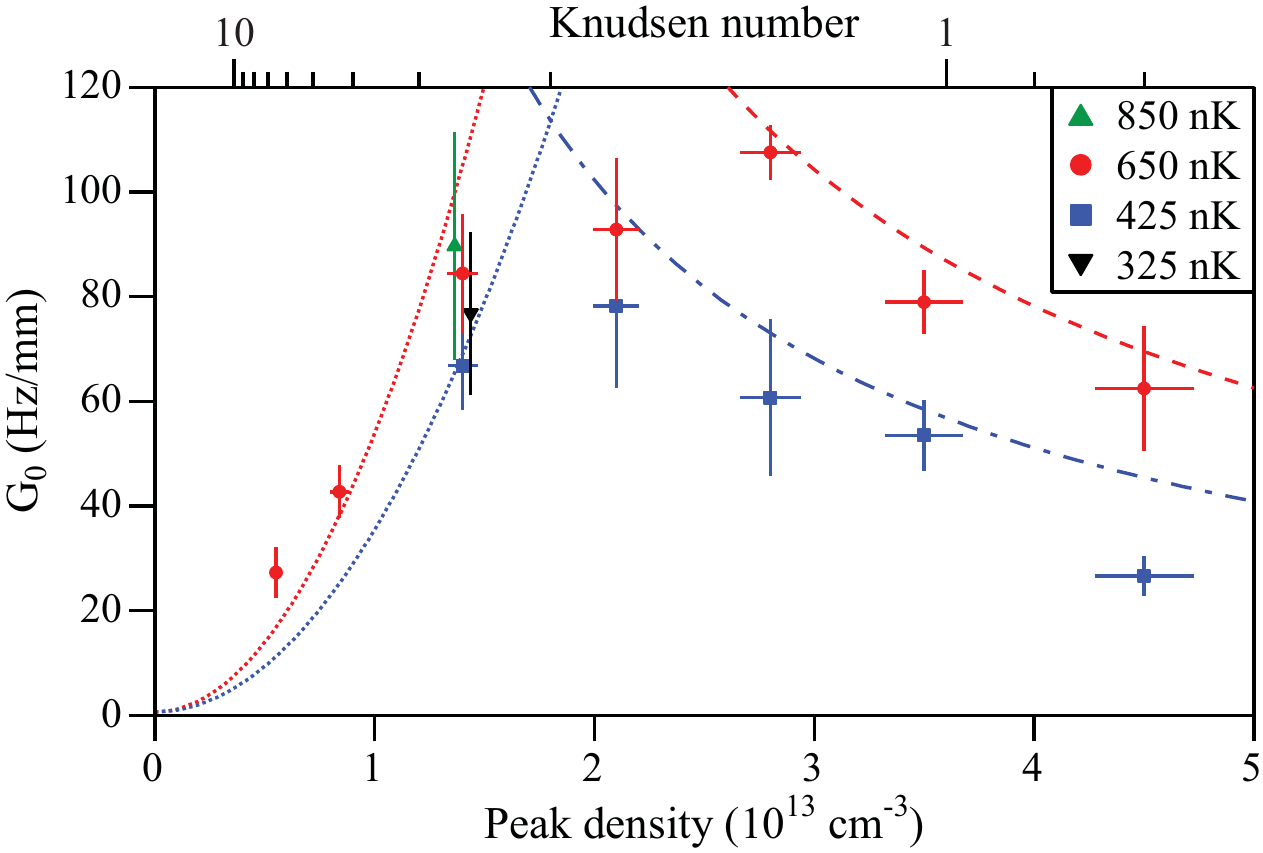}
    \caption{Stabilizing gradient $G_0$ as a function of peak density and temperature for $\lambda_0=73~\mu$m. Points at $n_0=1.4\times 10^{13}$~cm$^{-3}$ have been offset slightly for clarity. The data spans the crossover from collisionless to hydrodynamic regimes, as indicated by the Knudsen number for $\lambda_0=73~\mu$m. The data is described by $G\sub{hydro}$ at high density for $T=650$~nK (dashed) and 425~nK (dash-dot). Dotted lines show $G\sub{hydro}$ for $\lambda\sub{eq} = \ell\sub{3D}$ ($\mbox{Kn}=1$), demarking the region where the hydrodynamic model's prediction of a steady-state solution for some $\lambda_0$ is expected to be valid.}
    \label{Fig:GvsN}
\end{figure}

Density and temperature play important roles in the relationship between $\lambda\sub{eq}$ and $G_0$ (Fig.~\ref{Fig:GvsN}). At high density, the system nears the hydrodynamic regime and transport is inhibited; thus, smaller $G_0$ is needed to counteract spin gradients. Fig.~\ref{Fig:GvsN} shows the Knudsen number, $\mbox{Kn} = \ell/\lambda_0$, calculated at the domain-wall center using the mean free path $\ell$ and $\lambda_0=73~\mu$m. As Kn decreases to 1, though still in the crossover region between collisionless and hydrodynamic behavior, the $1/n$ dependence of $G\sub{hydro}$ agrees well with measured values of $G_0$. Since the atom cloud is collisionally thick at high $n_0$ and atoms are more localized, we use a radially averaged density in $G\sub{hydro}$, instead of an ensemble average. We note the $\lambda_0^{-3}$ dependence produces a 15\% uncertainty in $G\sub{hydro}$.

Furthermore, when temperature is lowered at high density, $G_0$ decreases. The quantitative agreement with $G\sub{hydro}$ is better at $T=650$~nK than at 425~nK, as assumptions contained within the model become more strained at lower temperature -- namely, that $\lambda_0\ll z_0$ and $\omega_z \tau \ll 1$. The former ensures slow, or unbounded, diffusion, while the latter is necessary for a large enough collision rate to ensure local equilibrium in the spin current. However, the approximation $\mu M\gg1$ strengthens at lower $T$ as collisions become more quantum in nature. Colder, smaller clouds do give larger uncertainties as the signal-to-noise ratio (SNR) drops; additionally, the highest density point at $T=425$~nK has $T/T_c=1.08$, and shot-to-shot fluctuations may occasionally include small condensates, which are inclined to phase separate to minimize energy in even small field gradients.

At low density, the hydrodynamic approximation breaks down, and $G_0$ deviates from $G\sub{hydro}$. The ultimate limits of the hydrodynamic model are shown by the dotted lines in Fig.~\ref{Fig:GvsN}, where $G\sub{hydro}$ is evaluated for the largest $\lambda\sub{eq}$ that satisfies $\mbox{Kn}=1$ for a given $n_0$ and $T$, i.e.~$\lambda\sub{eq}=\ell\sub{3D}$. Here $\ell\sub{3D}$ uses an ensemble-averaged density that is more appropriate in the low-density limit, instead of the radially averaged Kn shown on the Fig.~\ref{Fig:GvsN} axis. The hydrodynamic model predicts that steady-state domains exist for the parameter space to the right of the dotted lines, but does not guarantee the existence of any stable domains to the left. In fact, at low densities the measured value of the stabilizing gradient corresponds roughly to $\lambda\sub{eq}=\ell\sub{3D}$. A linearized moment-method analysis of Eq.~\ref{eq:Boltz} in the collisionless limit suggests that steady-state solutions do indeed exist for large domain walls ($\lambda_0>z_0$) at small gradients, but it is not clear that narrow domain-wall preparations such as initialized here can be stabilized in the collisionless regime.

Low density complicates the experimental technique of using domain-wall relaxation to find $G_0$. The second-lowest density in Fig.~\ref{Fig:GvsN}, $n_0=0.84\times10^{13}$~cm$^{-3}$, shows unambiguous signatures of steady-state domains and behavior consistent with higher values of $n_0$, but the $G_0$ found for $n_0=0.55\times10^{13}$~cm$^{-3}$ may not truly represent a stable domain solution for $\lambda_0 = 73~\mu$m. Large transients are observed here, where the initial spin current lies further from its equilibrium value, and rapid dephasing in the collisionless regime renders analysis of $\dot{\lambda}$ less reliable in determining $G_0$. It is likely that the measurement of $G_0$ at this density is tainted by transients and also represents an average value for $0<M<1$ due to dephasing. Indeed, analysis of $\dot{\lambda}$ for very short times ($t<8$~ms) and small gradients ($G<40$~Hz/mm) suggests that $G_0$ should be several times higher; however, measurements at higher $G$ reveal only rapid dephasing and even faster domain-wall relaxation rates. Thus it appears that steady-state solutions may not exist for narrow domain walls in the collisionless limit.

\begin{figure}[ht]
    \includegraphics[width=\linewidth,clip]{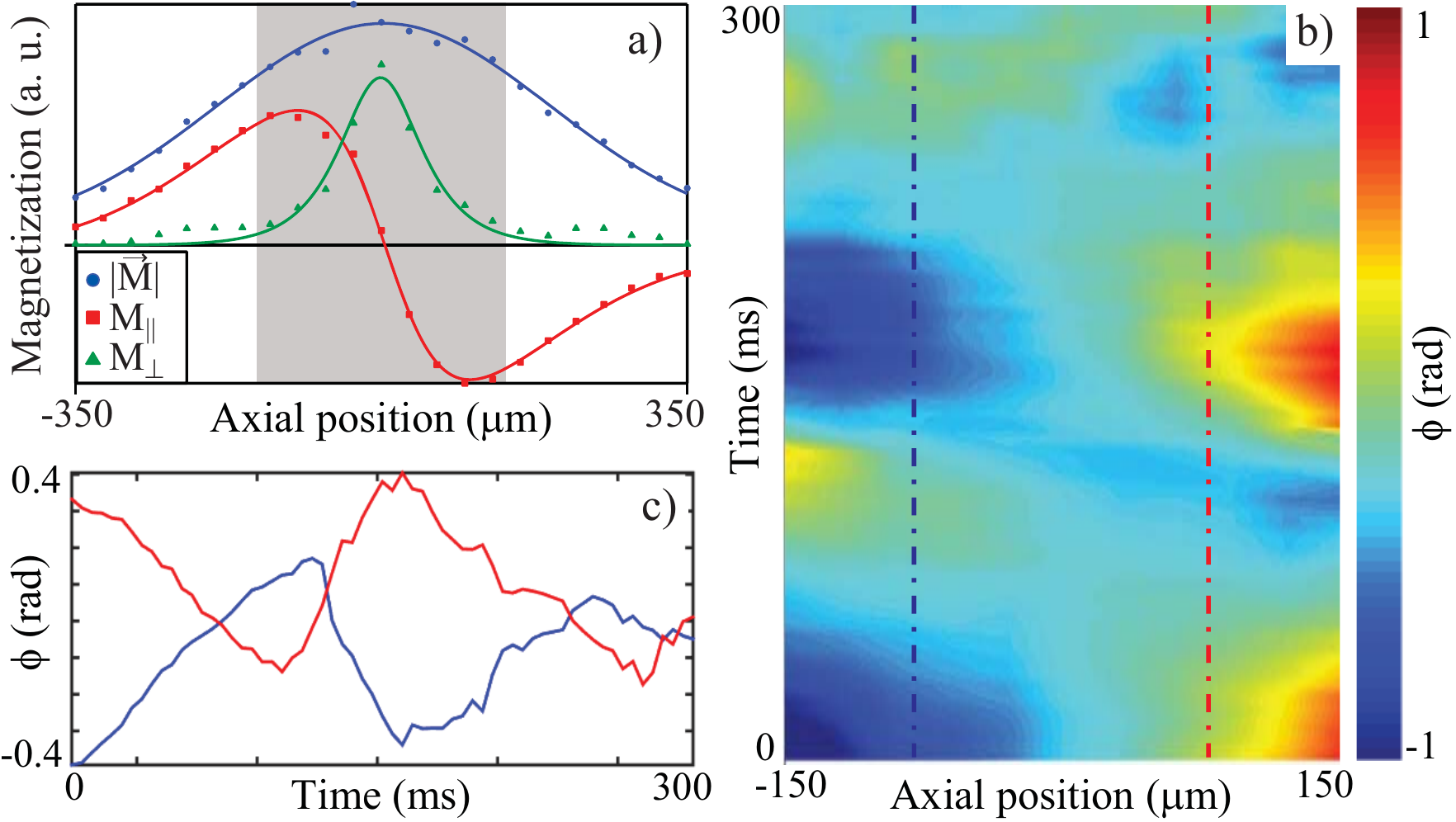}
    \caption{a) Initial transverse spin magnitude $M_\perp$ as measured from Ramsey fringe amplitudes. Also shown are $M_\parallel$ and $|\vec{M}|$. Minimal decoherence is expected during the short preparation sequence. Any discrepancies in $|\vec{M}| \simeq \sqrt{M_\parallel^2+M_\perp^2}$ are attributed to noise degrading Ramsey fringe fits. Solid lines are fits to a Gaussian ($|\vec{M}|$), and the same Gaussian multiplied by sech~$z/\lambda$ ($M_\perp$) and $\tanh z/\lambda$ ($M_\parallel$). b) Spatiotemporal evolution of the orientation of $\vec{M}_\perp(z,t)$, from the shaded region in a). c) The time evolution of $\phi$ exhibits dipolar spin wave oscillations, highlighted for two locations in the domain wall [dotted lines in (b)].}
    \label{Fig:transverse}
\end{figure}

Lastly, an important component to understanding the stabilization of domains is the behavior of the transverse magnetization $\vec{M}_\perp = M_\perp e^{i\phi}$. We study the dynamics of $\vec{M}_\perp$ for the conditions in Fig.~\ref{Fig:profiles}(b) by applying a $\pi/2$-pulse after a variable delay time to produce Ramsey fringes \cite{niroomand2015diffusion}. The amplitude and phase of Ramsey fringes give the magnitude $M_\perp$ and phase angle $\phi$ respectively. There are several striking features in the behavior of $\vec{M}_\perp$ (Fig.~\ref{Fig:transverse}). First, $\vec{M}_\perp$ undergoes phase oscillations, i.e. a dipolar transverse spin wave. This represents a decoupling in dynamics between a purely transverse spin wave and the stable longitudinal spin domains, as opposed to the coupled transverse-longitudinal spin dynamics observed in Ref.~\cite{niroomand2015diffusion} at $G=0$. The frequency of this spin wave approaches $\omega_z$, consistent with strongly driven, highly nonlinear dipolar spin waves \cite{mcguirk2010optical}. Secondly, stable longitudinal domains preclude any diffusion of transverse spin as well, and the transverse spin wave remains trapped within the domain wall.

A confined purely transverse spin wave is necessary to support the steady-state domains. The precessing transverse spin current provides the spin rotation as atoms move across the domain wall from $M_\parallel=1$ to -1. Though microscopic spins remain coupled via the ISRE, the macroscopic magnetization decouples completely. By contrast, in the absence of an effective field gradient, the longitudinal magnetization gradient induces spin currents that determine the transverse phase gradient, which would be stable with no field gradient. However, longitudinal domains are unstable at $G=0$ and undergo dipole oscillations, dragging the transverse phase with them and leading to rapid collapse and revival of $M_\perp$ \cite{niroomand2015diffusion}.

In conclusion, we have demonstrated stabilization of spin domains in a non-degenerate gas using optically-induced effective magnetic field gradients. The gradients allow domain lifetimes more than 30 times the classical diffusion time. This effect is driven by quantum symmetry in microscopic atom-atom interactions, and leads to decoupling of the dynamics of the longitudinal spin domains and a trapped transverse spin wave. Our experiments show good agreement with a hydrodynamic approximation at high density and suggest limitations to steady-state domains in the collisionless limit.

This project is supported by NSERC. S.D.~Graham and D.~Niroomand contributed equally to this work.

\bibliography{StableDomains}

\end{document}